\documentclass[a4paper]{jpconf}
\usepackage{graphicx}

\usepackage{amsfonts}
\usepackage{amsmath}
\usepackage{amssymb}
\usepackage{graphicx}

\providecommand{\U}[1]{\protect\rule{.1in}{.1in}}
\newcommand{\ba}{\begin{eqnarray}}
\newcommand{\ea}{\end{eqnarray}}
\def\beq{\begin{equation}}
\def\eeq{\end{equation}}

\begin{document}

\begin{flushright}
CERN-TH-PH/2009-034\\
March 2009
\end{flushright}

\title{Exploration of Possible Quantum Gravity Effects with Neutrinos II: Lorentz
Violation in Neutrino Propagation}

\author{Alexander Sakharov$^{1,2,3}$, John Ellis$^{1}$, Nicholas Harries$^{1}$,
Anselmo Meregaglia$^{4}$, Andr\' e Rubbia$^{2}$}

\address{$^{1}$ Theory Division, Physics Department, CERN, Geneva 23 CH 1211,
Switzerland\\
$^{2}$ Swiss Institute of Technology ETH-Z\"urich, 8093, Z\"urich,
Switzerland\\
$^{3}$ Department of Physics, Wayne State University, Detroit, MI 48202, USA\\
$^{4}$ IPHC, Universit, Louis Pasteur, CNRS/IN2P3, Strasbourg, France\\
}

\ead{Alexandre.Sakharov@cern.ch}

\begin{abstract}
It has been suggested that the interactions of energetic particles
with the foamy structure of space-time thought to be generated by
quantum-gravitational (QG) effects might violate Lorentz invariance,
so that they do not propagate at a universal speed of light.  We
consider the limits that may be set on a linear or quadratic
violation of Lorentz invariance in the propagation of energetic
neutrinos, $v/c = [1 \pm (E/M_{\nu QG1})]$ or $[1 \pm (E/M_{\nu
QG2})^2]$, using data from supernova explosions and the OPERA
long-baseline neutrino experiment.
\end{abstract}

\section{Introduction}

Neutrinos from astrophysical sources  and long-baseline experiments
are powerful probes of potential new physics. They have already been
used to discover and measure the novel phenomena of neutrino
oscillations, thereby establishing that neutrino have masses
\cite{Strumia:2006db, Totsuka:1991dm}. It has
been suggested that the space-time foam
due to QG fluctuations might cause energetic particles to propagate
at speeds different from the velocity of light, which would be
approached only by low-energy massless particles~\cite{mavrik1,foam,
gambini}. Any deviation from the velocity of light at high energies
might be either linear or quadratic, $\delta v/c = (E/M_{QG1})$ or
$(E/M_{QG2})^2$, and might be either subluminal or superluminal.
Such effects are, in principle, easily distinguishable from the
effects of neutrino masses, since they depend differently on the
energy $E$.

There have been many probes of such Lorentz-violating effects on
photon propagation from distant astrophysical objects such as
gamma-ray bursters~\cite{amellis}, pulsars~\cite{pulsar} and active
galactic nuclei~\cite{Albert:2007qk}. These tests have looked for
delays in the arrival times of energetic photons relative to
low-energy photons, and their sensitivities improve with the
distance of the source, the energies of the photons, the accuracy
with which the arrival times of photons can be measured, and the
fineness of the time structure of emissions at the astrophysical
source. The sensitivities of these tests have reached $M_{\gamma QG1}
\sim 10^{18}$~GeV and $M_{\gamma QG2} \sim 4 \times
10^{10}$~GeV for linear and quadratic violations of Lorentz
invariance, respectively~\cite{Combine}.

At least one QG model of space-time foam~\cite{equiv,refract_last}
suggests that Lorentz violation (LV) should be present only for particles
without conserved internal quantum numbers, such as photons, and
should be absent for particles with electric charges, such as
electrons~\cite{mavrik1}. Indeed, astrophysical data have been used to set very
stringent limits on any LV in electron propagation.
However, these arguments do not apply to neutrinos, since they are
known to oscillate, implying that lepton flavour quantum numbers are
not conserved. Moreover, neutrinos are often thought to be Majorana
particles, implying that the overall lepton number is also not
conserved, in which case QG effects might also be present in
neutrino propagation~\cite{LeptonNumber,volkov}. It is therefore 
interesting to study experimentally the possibility of Lorentz
violation in neutrino propagation~\cite{volkov,lvrub}.

Experimental probes of LV in neutrino propagation are
hindered by the relative paucity of neutrino data from distant
astrophysical sources, and require the observation of narrow time
structures in neutrino emissions. However, there has been one
pioneering experimental study of possible LV using
the long-baseline MINOS experiment exposed to the NuMI neutrino beam
from Fermilab, which found a range of neutrino velocities $-2.4
\times 10^{-5} < (v - c)/c < 12.6 \times 10^{-5}$ allowed at the
99\% C.L.~\cite{Rebel:2008th}. Assuming an average neutrino energy
of 3~GeV, and allowing for either linear or quadratic Lorentz
violation: $v/c = [1 \pm (E/M_{\nu QG1})]$ or $[1 \pm (E/M_{\nu
QG2})^2]$, the MINOS result~\cite{Rebel:2008th} corresponds in the
case of linear LV to $M_{\nu QG1}
> 1 (4) \times 10^5$~GeV for subluminal (superluminal)
propagation, and in the case of quadratic LV to
$M_{\nu QG2} > 600 (250)$~GeV.

In this report we describe limits on LV established
in~\cite{lvrub} by using
neutrino supernova 1987a data from the Kamioka II
(KII) \cite{k2sn1987a}, Irvine-Michigan-Brookhaven (IMB)
\cite{imbsn1987a} and Baksan detectors \cite{baksan1987a}. 
We find limits that are significantly more stringent than those established
using the
MINOS detector. We also assess the improved sensitivity to Lorentz
violation that could be obtained if a galactic supernova at a
distance of 10~kpc is observed using the Super-Kamiokande (SK) detector.

We then discuss the sensitivity to LV of the OPERA
experiment at the CNGS neutrino beam from
CERN. We point that substantial improvements in sensitivity of CNGS to LV in
neutrino probe would result if one could
exploit the RF bucket structure of the spill for neutrino events occurring in
the rock upstream from OPERA. In this case, the sensitivity that could be achieved
for quadratic LV is better than that obtained
from supernova 1987a, and even improves on the sensitivity possible
with a future galactic supernova.

\section{Supernovae data analysis}
%
In this Section we discus the ability of supernova
data to
test LV. In particular, we analyze
the data from the supernova SN1987a,
the first supernova from which neutrinos have been detected, giving
bounds at the $95\%$ C.L.. Then we simulate a possible future
galactic supernova and discuss the potential of the next generation
of neutrino detectors, represented by Super-Kamiokande (SK), to
improve this bound.

The detection of neutrinos from SN1987a in the Large Magellanic
Cloud (LMC) remains a landmark in neutrino physics and astrophysics.
Although only a handful of neutrinos were detected by the
KII~\cite{k2sn1987a}, IMB~\cite{imbsn1987a} and
Baksan~\cite{baksan1987a} detectors,
they provided direct evidence of the mechanism by which a star
collapses, and the role played by neutrinos in this mechanism
\cite{Totsuka:1991dm}. The numbers and energies of the neutrinos
observed were consistent with the expected supernova energy release
of a few times $10^{53}$~ergs via neutrinos with typical energies of
tens of MeV. A future galactic supernova is expected to generate up
to tens of thousands of events in a water-{\v C}erenkov detector
such as SK, which will clarify further theories of the supernova
mechanism and of particle physics~\cite{Ikeda:2007sa}.

We are interested in the possibility of QG
effects leading to LV modifications to the propagation of
energetic particles, and hence to dispersive effects, specifically a
non-trivial refractive index. These dispersive properties of the
vacuum would lead to an energy dependence in the arrival times of
neutrinos. Therefore, any data set comprising 
both the time and energy of each neutrino event can be
analyzed by inverting the dispersion that would be caused by any
hypothesized QG effect. The preferred value of any energy-dependence parameter
would minimize the duration (time spread) of the supernova neutrino signal.

Assuming either a linear or a quadratic
form of LV: $v/c = [1 \pm (E/M_{\nu QG1})]$ or $[1
\pm (E/M_{\nu QG2})^2]$, a lower limit on $M_{\nu QG1}$ and $M_{\nu
QG2}$ may be obtained by requiring that the emission peak not be
broadened significantly. A non-zero value of $M_{\nu QG1}^{-1}$ or
$M_{\nu QG2}^{-1}$ might be indicated if it reduced significantly
the duration (time spread) of the neutrino signal. The duration
(time spread) of the neutrino signal can be quantified using different
estimators depending on the amount of available statistics and time
profile of the data set, if applicable~\footnote{ Statistically poor event
lists,
such as that for SN1987a, the only one currently available in supernova
neutrino 
astronomy, do not allow the time profile to be classified, because time binning
is
impractical and one cannot apply nonparametric statistical tests to
unbinned data.}. In the following, we outline two estimators for analyzing
neutrino signals (see~\cite{lvrub} for details), that we use first to
quantify the limits obtainable from the SN1987a neutrino data and then
the sensitivities that would be provided by a possible future
galactic supernova signal.

{\it Minimal Dispersion (MD) Method.}  We assume that the data set
consists of a list of neutrino events
with measured energies $E$ and arrival times $t$ (for
details, see~\cite{lvrub} and references therein). 
In the first method, we consider event lists
with a relatively low number of events, that do not allow a
reasonable time profile to be extracted. In this case we consider
the time
dispersion of the data set, quantified by
$\sigma_{t}^{2}\equiv\langle\left(t-\langle
t\rangle\right)^{2}\rangle$,
where $t$ is the time of each detected event. We then apply an
energy-dependent time shift $\Delta t=\tau_{l} E^{l}$, where
$\tau_{l}=L/c M_{\nu QGl}^{l}$, varying $M_{\nu QGl}$ so as remove
any assumed dispersive effects. The `correct' value of the time shift $\tau_{l}$
should always compress the arrival times of the neutrino
events. Any other (`uncorrect') value of $\tau_{l}$ would spread the
events in time, relative to the `correct' value. We denote by
$\tau_{l}^{min}$ the value that minimizes the spread in the arrival times.
In order to estimate the uncertainties in $\tau_{l}^{min}$, we
use a Monte Carlo simulation to repeat the calculation of
$\tau_{l}^{min}$ including the energy and statistical uncertainties.
We then make a Gaussian fit and use it to quote best-fit parameters
and errors.

{\it Energy Cost Function (ECF) Method.} This is a different analysis
technique that is mostly applicable to event
lists
that are statistically rich. This means
that one can combine the neutrino events into a time profile exhibiting
pulse features that can be distinguished 
from a uniform distribution at high confidence level. For the analysis we first
choose the most
active (transient) part of the signal $(t_{1};t_{2})$~\footnote{ The
most active part of the signal
can be chosen by fitting the binned time profle or using a Kolmogorov-Smirnov (KS)
statistic~\cite{lvrub}.
In the case of a multipulse structure  of the time profile, several windows may
be analized separately.}. { Having chosen this window, we scan over its whole
support the time distribution
of all events, shifted by $\Delta t=\tau_{l} E^{l}$, and sum the energies 
of events in the
window. This procedure is repeated for many values of $\tau_l$,
chosen so that the shifts $\Delta t$ match the precision of the arrival-time
measurements, thus defining the `energy cost function' (ECF).
The maximum of the ECF indicates the value of $\tau_l$ that best  recovers
the signal, in the sense of maximizing its power (amount of energy in a
window of a given time width $t_{2}-t_{1}$). This
procedure is then repeated for many Monte-Carlo (MC)
data samples generated by
applying to the measured neutrino energies the estimated Gaussian
errors. 

Neutrinos from SN1987a were detected in three detectors, KII~\cite{k2sn1987a}, 
IMB~\cite{imbsn1987a} and Baksan~\cite{baksan1987a}, and 
the times and energies of the events are given in~\cite{lvrub}. We
calculated the
minimum dispersion 1000 times for each
data set, so as to include the smearing from uncertainties~\cite{lvrub}.
This analysis constrains the scale at which LV may enter the neutrino sector
to be $M_{\nu QG1} > 2.7 \times 10^{10}~{\rm GeV~or}~M_{\nu QG1}
> 2.5 \times 10^{10}~{\rm GeV}$ at the $95\%$ C.L. for the linear subluminal and
superluminal models
respectively. The corresponding limits for the quadratic models are
$M_{\nu QG2} > 4.6 \times 10^{4}~{\rm GeV~or}~M_{\nu QG2} > 4.1
\times 10^{4}~{\rm GeV}$
at the $95\%$ C.L. for the subluminal and superluminal versions,
respectively.

The detection of a galactic supernova would provide improved
sensitivity to the scale at which LV might enter the
neutrino sector, due to an increase in the number of neutrinos which
would be detected. The number of events would also increase because
the current neutrino detectors are larger than those used to detect
neutrinos from SN1987a. However, these effects would be partially
offset because $\tau_{l}\propto L$ and therefore the time-energy
shift will be reduced if, as expected, the supernova takes place
within the galactic disc at a distance $\sim 10$~kpc, compared to
SN1987a in the LMC at a distance of $\sim 51$~kpc. The closer
distance would also increase
the number of neutrinos that are expected to be detected,
compared to SN1987a. For definiteness, we use here a Monte Carlo simulation
of the Super-Kamiokande (SK) neutrino detector, but note that other
neutrino detectors could also probe this
physics~\cite{NeutrinoDetectors}. Simulations estimate that the
number of events detected in SK from a supernova at 10~kpc would be
of the order of 10,000 \cite{Ikeda:2007sa}. In order to analyze at
what scales LV could be probed by the detection of
galactic supernova neutrinos, we made Monte Carlo simulations with
various levels of linear and quadratic LV. We used
the energy spectra of neutrinos from the Livermore
simulation~\cite{Totani:1997vj}, which is shown in Fig.~\ref{fig:SNSpectrum}, and
the detector properties given
in~\cite{Tomas:2003xn}.
\begin{figure}[h]
\begin{minipage}{14pc}
\includegraphics[width=14pc]{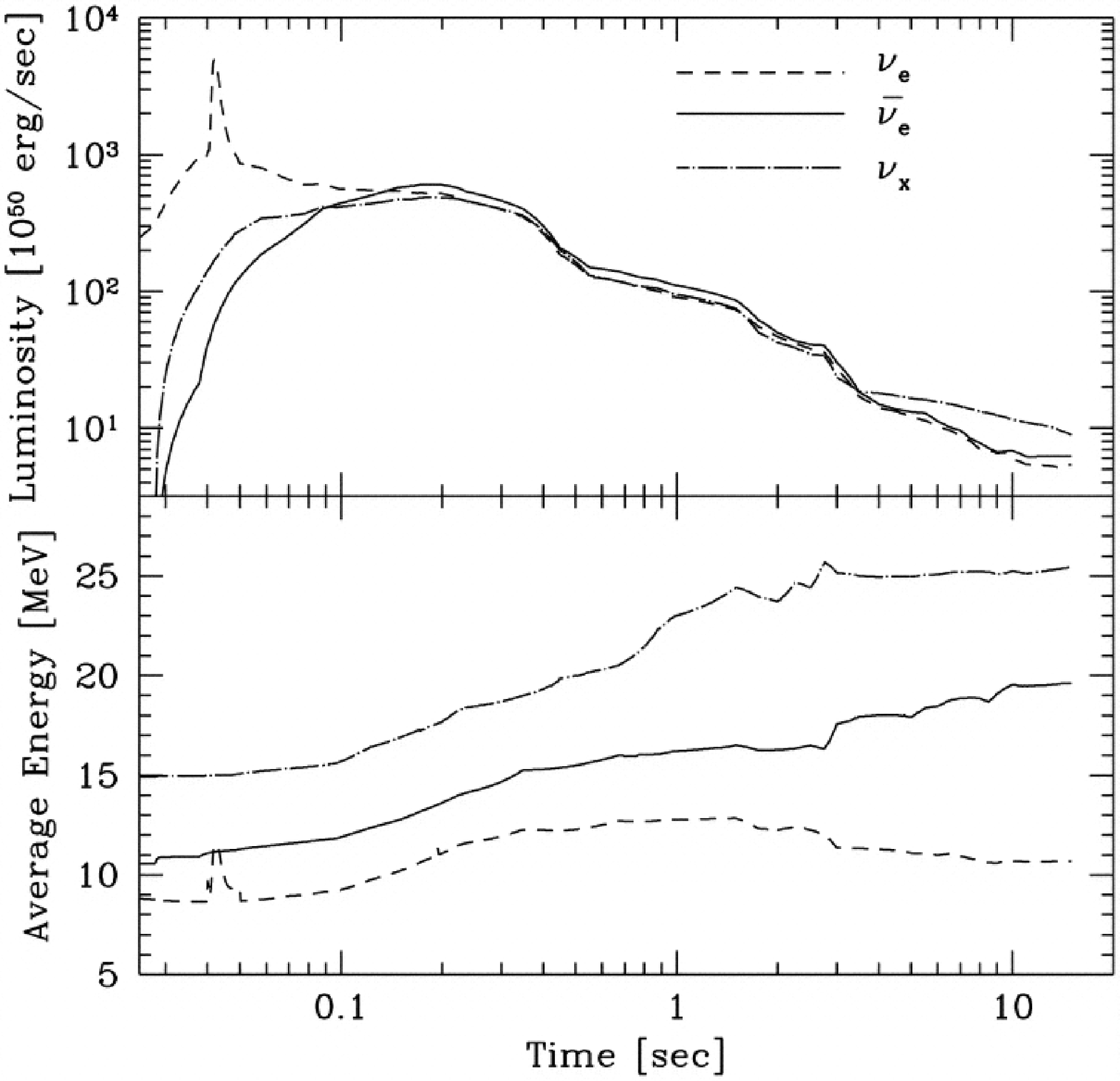}
\caption{\label{fig:SNSpectrum}The neutrino energy spectra from the Livermore
simulation~\cite{Totani:1997vj}.}
\end{minipage}\hspace{2pc}%
\begin{minipage}{14pc}
\includegraphics[width=18pc]{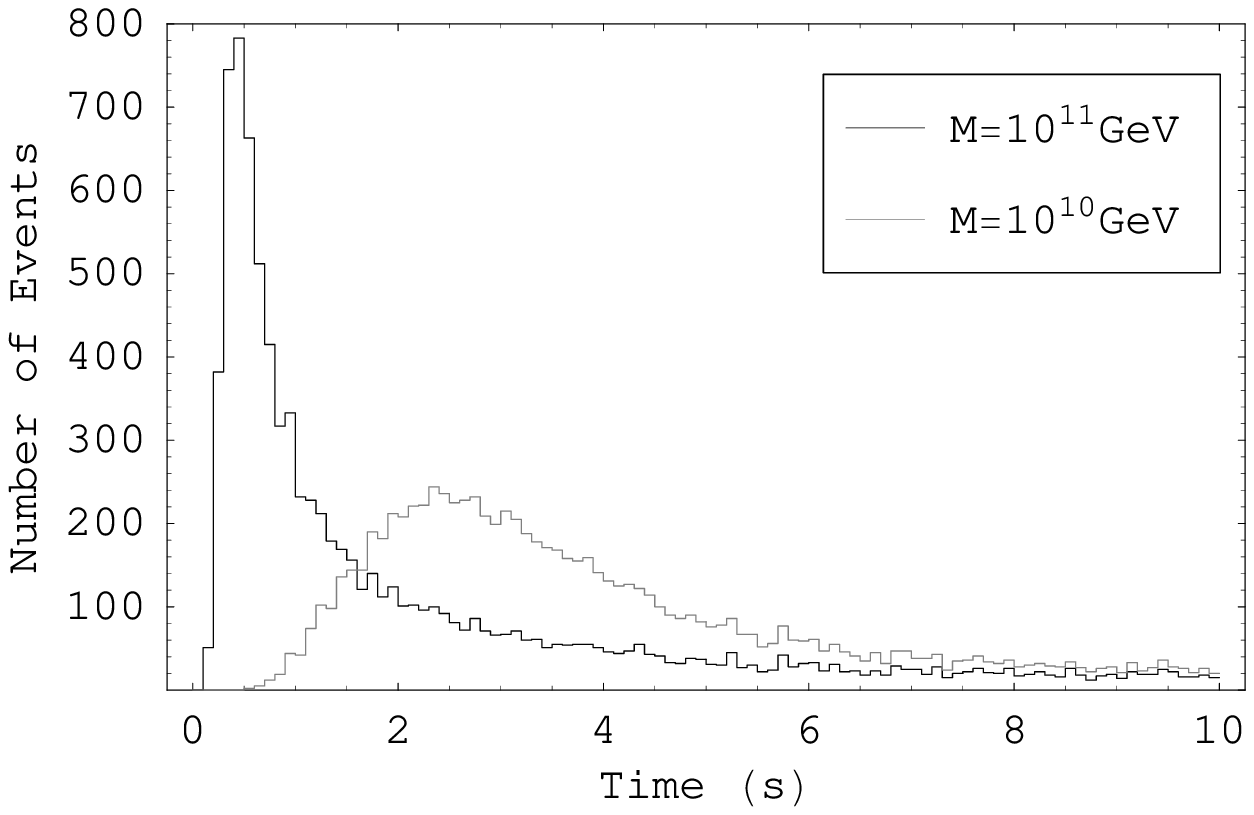}
\caption{\label{fig:events}The time distribution of events predicted by our
Monte
Carlo simulation for the case of subluminal LV at the
mass scales $M=10^{10}GeV$ and $M=10^{11}GeV$.}
\end{minipage} 
\end{figure}
We show in Fig.~\ref{fig:events} results from our Monte Carlo
simulation including both charged-current and neutral-current events
for linear sublminal LV at the energy scales $M_{\nu
QG1}=10^{10}GeV$ and $M_{\nu QG1}=10^{11}GeV$, including
oscillations corresponding to the normal hierarchy and assuming that
the atmospheric resonance is adiabatic. The signal has spread out
and shifted in time, as we would expect. 

We have applied the MD and the maximal ECF methods with various
energy weightings to the Monte Carlo data with $M_{QG1}=10^{10}$~GeV
in order to estimate the level of LV. In this way, we established
that data
from a future galactic supernova could place strong $95\%$ C.L.
limits on the range of $M_{\nu QG1}$ if it is lower than
$10^{11}$~GeV. In the limit of negligible LV ($M_{\nu
QG1} \ge 10^{12}$~GeV), we find the lower limits $M_{\nu QG1}> 2.2
\times 10^{11}GeV$ and $M_{\nu QG1}> 4.2 \times 10^{11}GeV$ at the
$95\%$ C.L. for subluminal and superluminal models, respectively. In the
case
of large $M_{\nu QG2}$, we find the lower limits $M_{\nu QG2}> 2.3
\times 10^{5}$~GeV and $M_{\nu QG2}> 3.9 \times 10^{5}$~GeV at the
$95\%$ C.L. for subluminal and superluminal models, respectively, in
the quadratic case.

\section{CNGS and the OPERA Experiment}

In this Section we discuss the sensitivities to LV in
neutrino propagation that could be provided by the OPERA experiment
in the CNGS neutrino beam. 

The energy spectrum of the calculated CNGS $\nu_\mu$ flux is
reproduced in Fig.~\ref{fig:spectrum}. Its average neutrino energy
is $\sim 17$~GeV, significantly higher than that of the NuMI beam.
Since the CNGS baseline is almost identical with that the NuMI beam,
this gives some advantage to OPERA, assuming that it can attain
similar or better timing properties. We recall that the CNGS
beam is produced by extracting the SPS beam during spills of length
$10.5~\mu$s (10500~ns). Within each spill, the beam is extracted in
2100 bunches separated by 5~ns. Each individual spill has a
$4-\sigma$ duration of 2~ns, corresponding to a Gaussian RMS width
of 0.25~ns \cite{Meddahi:2007zz}.

\begin{figure}[h]
\begin{minipage}{14pc}
\includegraphics[width=16pc]{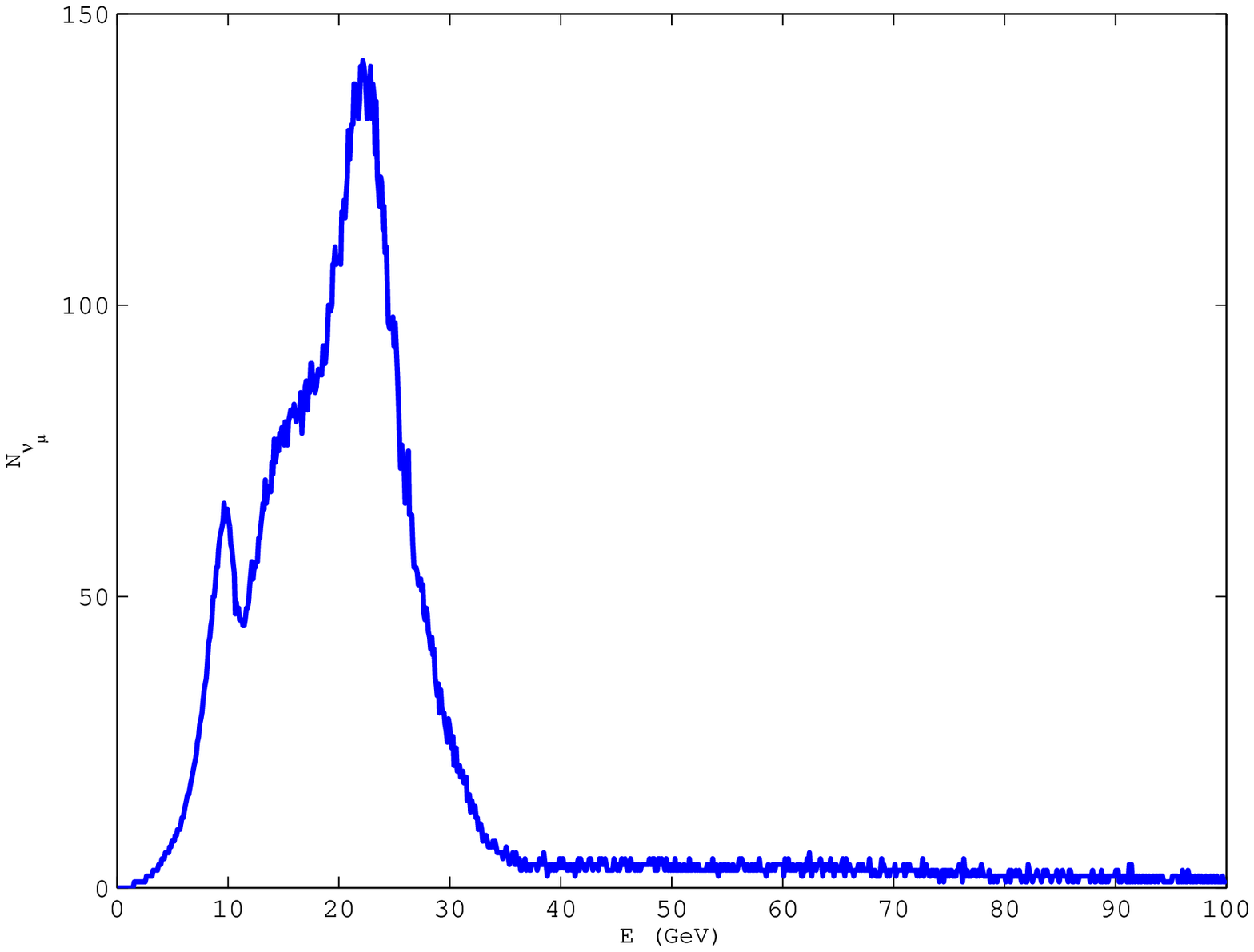}
\caption{\label{fig:spectrum}The expected CNGS neutrino beam energy
spectrum~\cite{Meddahi:2007zz}.}
\end{minipage}\hspace{2pc}%
\begin{minipage}{14pc}
\includegraphics[width=18pc]{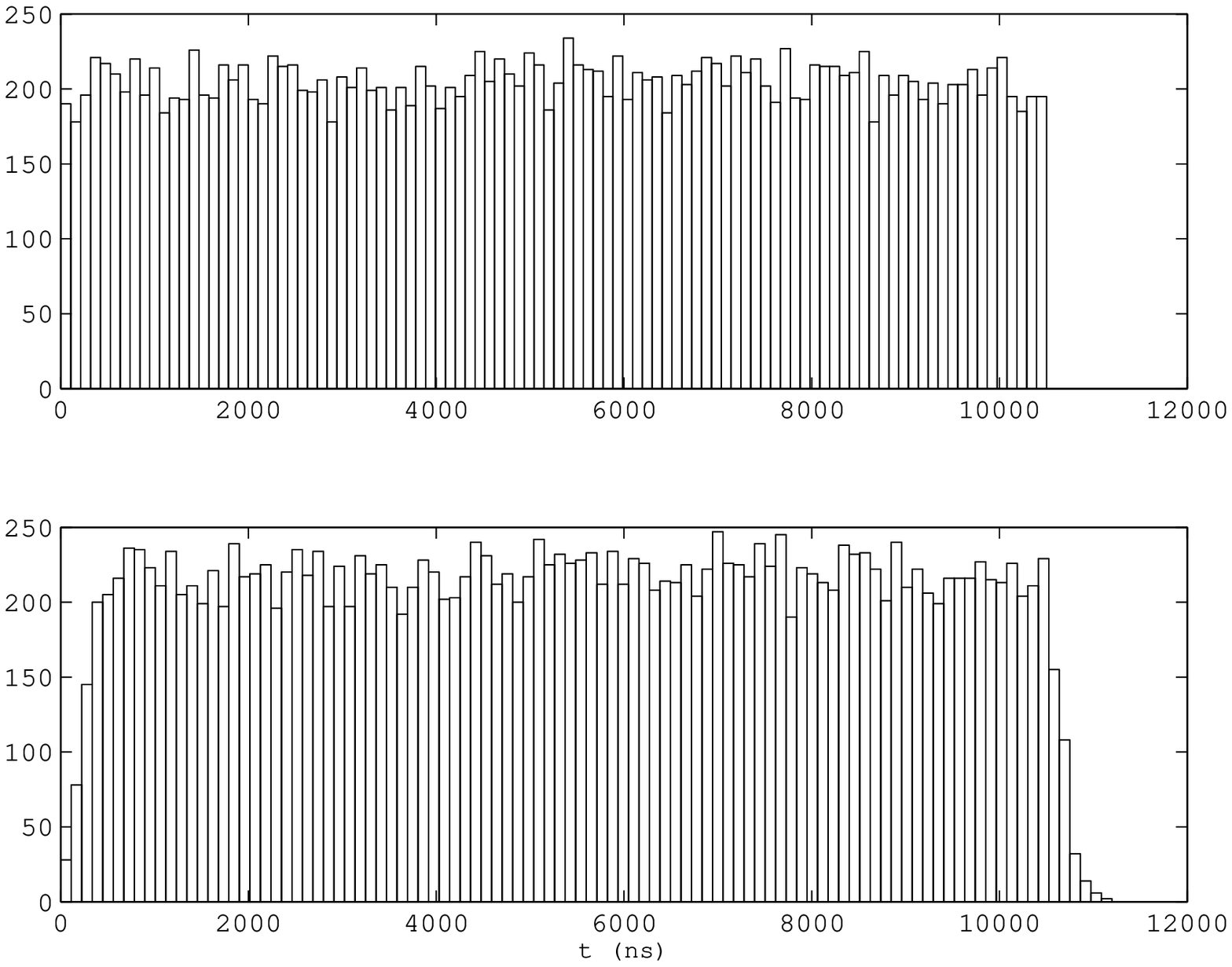}
\caption{\label{fig:smear}The time structure of events in the CNGS beam,
without LV (upper panel), and with time delay at
the level
of $\tau =5$~ns/GeV (lower panel).}
\end{minipage} 
\end{figure}

We introduce a `slicing estimator'~\cite{lvrub}, based on the fact that if some
energy-dependent time delay is encoded into the time structure of
the spill by propagation of the neutrinos before detection, one
should observe a systematic increase in the overall time delay of
events as their energies grow. Therefore, we propose cutting the
energy spectrum of the neutrino beam into a number of energy slices,
and searching for a systematic delay in the mean arrival times of
the events belonging to different energy slices that increases with
the average energy of the slice.

In order to illustrate this idea, we perform a simple exercise
simulating the sensitivity of the slicing estimator for a time delay
depending linearly on the neutrino energy: $\Delta t=\tau E$,
assuming $\approx 2 \times 10^4$ charged-current events, as are
expected to be observed in the 1.8 kton OPERA detector over 5 years of
exposure time to the CNGS beam. We envisage superposing all the CNGS
spills with a relative timing error $\delta t$. Since each spill has
2100 bunches, we expect about 10 events on average due to each set
of superposed bunches. As a starting-point, before incorporating the
relative timing error, the timing of each event has been smeared
using a Gaussian distribution with standard deviation 0.25~ns,
reflecting the bunch spread. We also incorporate the uncertainty in the relative
timing of the
bunch extraction and the detection of an event in the detector. The
overall uncertainty has three components: an uncertainty in the
extraction time relative to a standard clock at CERN, an uncertainty
in the relative timing of clocks at CERN and the LNGS provided by
the GPS system, and the uncertainty in the detector timing relative
to a standard clock in the LNGS. With the current beam
instrumentation, implementation of GPS and detector resolution, it
is expected that this will be similar to that achieved by MINOS in
the NuMI beam, namely $\sim 100$~ns. Such a timing error renders
essentially invisible the internal bunch structure of the CNGS
spill, which looks indistinguishable from a uniform distribution
generated with the same statistics, as shown in the upper panel of
Fig.~\ref{fig:smear}.

We next demonstrate in the lower panel of Fig.~\ref{fig:smear} the
effect of a time delay during neutrino propagation at the level of
$\tau_l =5$~ns/GeV, as would occur if $M_{\nu QG1} = 4.8\times 10^5\ {\rm GeV}$.
This would correspond to  a total delay $\sim 100$~ns at
the average energy of the CNGS neutrino beam. We smear the events with an energy
resolution of 20\%, and then cut
the sample into slices of about 1000 events each with increasing
energies. 

By making many realizations of the event sample with the Gaussian
$\delta t = 100$~ns smearing, one can understand the significance of
the shifts in the mean positions of the slices. Fig.~\ref{fig:shift}
shows the energy dependence of the shifts in the mean timings of the
slices of 1000 events with a delay $\tau_l =5$~ns/GeV encoded. These
points may be fitted to a straight line $\Delta\langle t\rangle =\tau_l\langle
E\rangle +b$.

\begin{figure}[h]
\begin{minipage}{14pc}
\includegraphics[width=14pc]{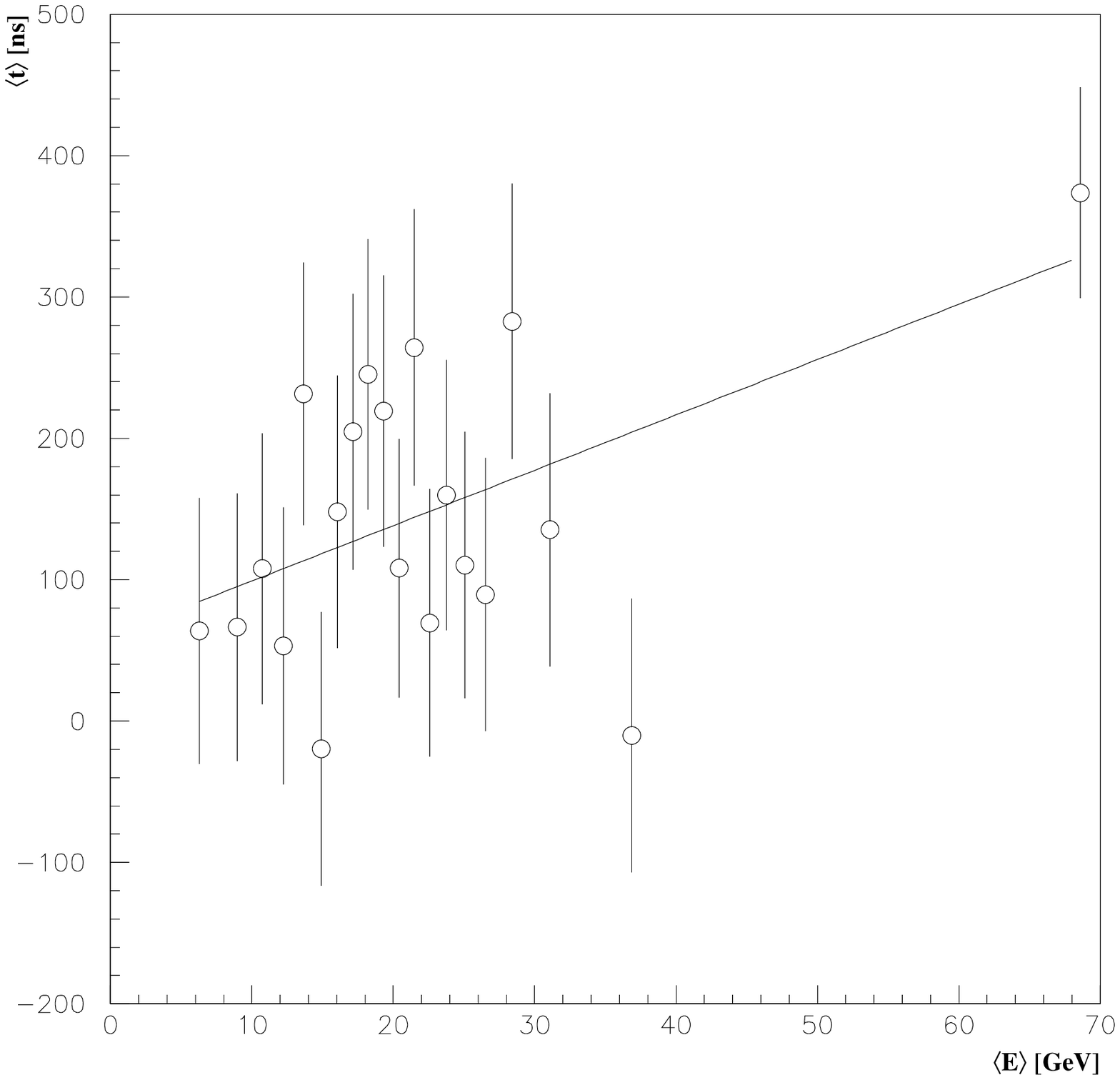}
\caption{\label{fig:shift}The measured shifts in the average arrival times of
neutrinos in 1000-event slices with increasing energies, assuming a
time delay during neutrino propagation at the level of $\tau
=5$~ns/GeV.}
\end{minipage}\hspace{2pc}%
\begin{minipage}{14pc}
\includegraphics[width=14pc]{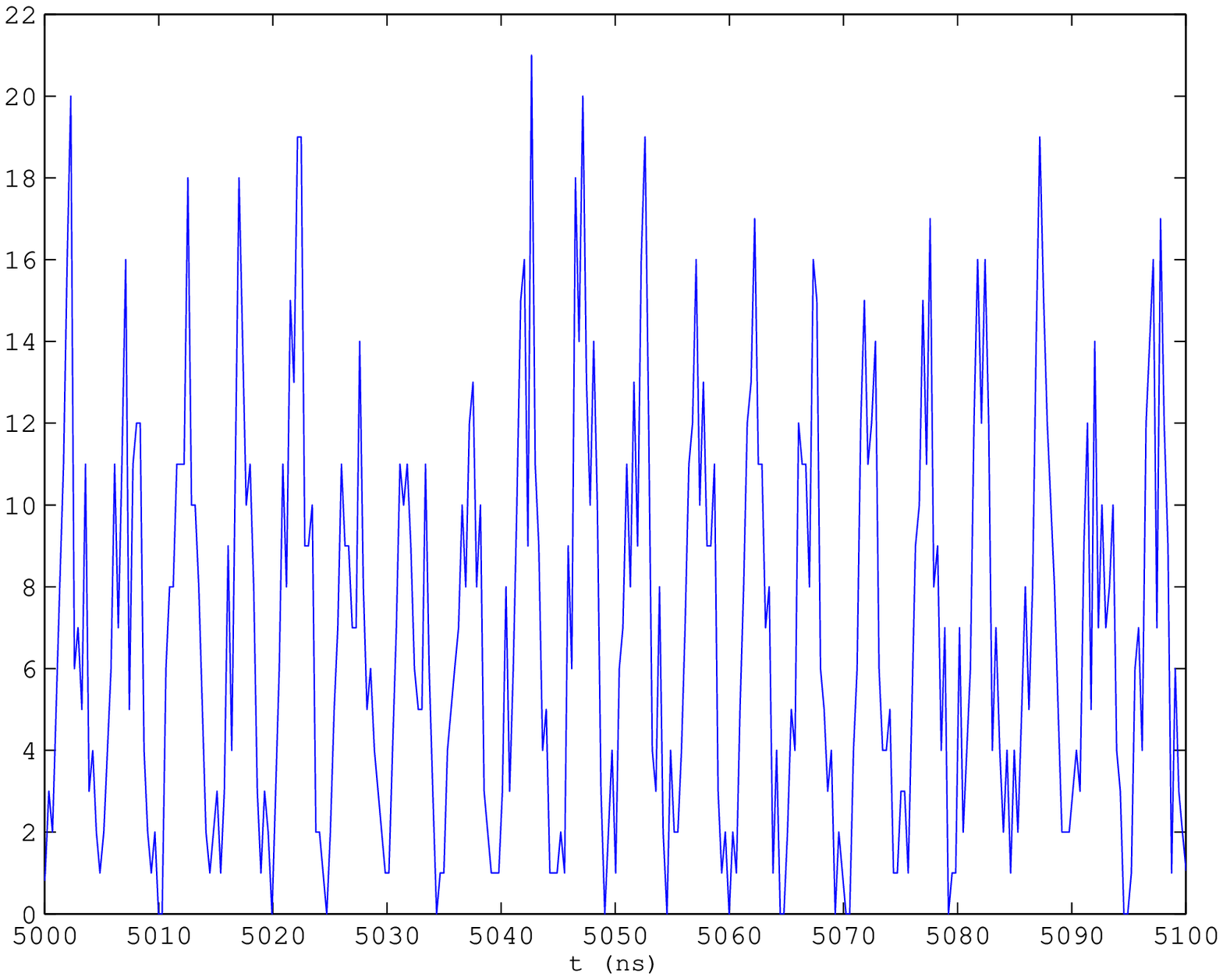}
\caption{\label{fig:batch_smeared1} A simulated realization of the bunch
structure for rock
events, incorporating a timing uncertainty $\approx 1$~ns. The
histogram is binned with a resolution suitable for resolving the
bunch structure.}
\end{minipage} 
\end{figure}
One obtains $\tau_{\rm
l95\%}=4.9(2.6)\ {\rm ns/GeV}$ at the 95\% C.L. for the subluminal
(superluminal) propagation schemes, corresponding to values of
the linear Lorentz-violating scale $M_{\nu QG1} =
4.9(9.2)\times 10^5\ {\rm GeV}$, yielding a
mean sensitivity to
$M_{\nu QG1}\simeq 7\times 10^5\ {\rm GeV}$. If the velocity of the neutrino
depends quadratically on the energy
of the neutrino, the slices should obey a parabolic fit $\label{quadrfit}
\Delta\langle t\rangle =\tau_q\langle E\rangle^2
+c$. In quadratic case we obtain the sensitivity 
$M_{\nu QG1}= 6.2(11)\times 10^3\ {\rm GeV}\simeq 8\times 10^3\ {\rm
GeV}$.
The stability of the slicing estimator has been checked by various methods
(see~\cite{lvrub} for the details) including spill edges fitting used in MINOS
analysis~\cite{Rebel:2008th}.

We recall that the OPERA detector may also be used to measure the
arrival times of muons from $2 \times 10^5$ neutrino events in the
rock upstream of the detector. Information on the neutrino energy is
missing in this measurement. Nevertheless, one can use methods that compare
overall the time
shift of the simulated data to the measured time distribution of the
rock events. In this spirit, applying to the $2\times 10^5$ expected
rock events the edge-fitting procedure described in~\cite{lvrub,Rebel:2008th},
we find a sensitivity to $M_{\nu QG1} \approx 2.4\times
10^6\ {\rm GeV}$, about three times better than previously, in the
case of linear energy dependence, and the same level of
sensitivity for the quadratic energy dependence.

We also explored the additional sensitivity that OPERA could obtain if
it could achieve a correlation between the SPS RF bunch structure
and the detector at the nanosecond level. Possible techniques for doing this are
outlined in details in~\cite{lvrub}. In Fig.~\ref{fig:batch_smeared1} we present
one particular
realization of a sample of simulated events which incorporates a
relative timing error of 1ns. Although the periodic bunch structure survives,
the signal itself
represents a time series with a relatively low signal-to-noise
ratio. The latter implies that the proper deconvolution to extract
isolated features cannot be made. In the other words, there is a
problem in fitting the fine structure of the signal with an
analytical function. Such a situation has been widely investigated
in analyses of the temporal profiles of gamma-ray bursters
(GRBs)~\cite{ccf_grb}. We therefore apply a cross correlation
function (CCF) method similar to that described in~\cite{ccf_grb}
but differing only in details of its adaptation~\cite{lvrub}. Namely, 
in~\cite{lvrub} we introduce a CCF for the temporal correlation of two time
series $A(t)$ and
$B(t+\tau_{l(q)})$ where $A(t)$ is a Monte Carlo simulation of the events with
no
dispersion effects, and $B(t+\tau_{l(q)})$ is the simulated data which
has the time shift required to invert the effect of the
energy-dependent dispersion. We average
over
several Monte Carlo simulations to include the statistical
uncertainties as well as performing time and energy smearing due to
the uncertainty in these measurements. We then calculate ${\rm CCF(\tau_{l(q)})}$
as a function of
$\tau_{l(q)}$ and find its maximum value. The value of $\tau_{l(q)}$ which
maximizes the CCF is an estimate of the true value of $\tau_{l(q)}$. To
find this estimate we fit a Gaussian to the peak of the resulting CCF, and
deduce the sensitivity of the CCF from the precision of
the position of the maximum for the Gaussian fit. In the case of linear
energy dispersion, the sensitivity obtained in~\cite{lvrub} corresponds to
 $M_{\nu QG1} \approx 6.6\times 10^7\ {\rm GeV}$. For the subluminal case, one
obtains $M_{\nu QG1} \approx 2.4\times 10^7\ {\rm GeV}$. The same CCF procedure
may also
be applied to the quadratic case~\cite{lvrub}. The limits deduced in this
case are $M_{\nu
QG2}= 3.6(4.9) \times 10^{4}\ {\rm GeV}\simeq 4\times 10^{4}\ {\rm
GeV}$ for superluminal (subluminal) propagation models.

The CCF calculated for the rock events gives the following sensitivity
lavels $M_{\nu
QG1}= 4.3(3.2)\times 10^8\ {\rm GeV}\simeq 4\times 10^8\ {\rm GeV}$
for the linear case, and $M_{\nu QG2}= 8.8(4.3) \times 10^{5}\ {\rm
GeV}\simeq 7 \times 10^{5}\ {\rm GeV}$ for the quadratic case. The
sensitivity in the quadratic case is significantly better than the
sensitivity estimated for a possible future galactic supernova.

\section{Conclusions}

We find from the SN1987a data lower limits on the scale of linear
LV in the neutrino sector that are $M_{\nu QG1}
> 2.68 \times 10^{10}$~GeV and $M_{\nu QG1}
> 2.51 \times 10^{10}$~GeV at the $95\%$ C.L. in the subluminal and
superluminal cases respectively. The corresponding limits for the
quadratic model are $M_{\nu QG2} > 4.62 \times 10^{4}$~GeV and
$M_{\nu QG2} > 4.13 \times 10^{4}$~GeV at the $95\%$ C.L. in the
subluminal and superluminal cases, respectively. We have also used a
Monte Carlo simulation of a galactic supernova at 10~kpc to estimate
how accurately LV could be probed in the future. We have
shown that it would be possible to place limits up to $M_{\nu QG1}> 2.2 \times
10^{11}$~GeV and $M_{\nu QG1}> 4.2 \times 10^{11}$~GeV at the $95\%$
C.L. for the subluminal and superluminal cases, respectively, for
linear models of LV, and $M_{\nu QG2}> 2.3 \times
10^{5}$~GeV and $M_{\nu QG2}> 3.9 \times 10^{5}$~GeV at the $95\%$
C.L. for the subluminal and superluminal cases, respectively, for
quadratic models of LV.

We find that,
using standard clock synchronization techniques,  the sensitivity of
the OPERA experiment would reach $M_{\nu QG1} \sim 7 \times
10^{5}$~GeV ($M_{\nu QG2} \sim 8 \times 10^{3}$~GeV) after 5 years
of nominal running. If the time structure of the SPS RF bunches
within the extracted CNGS spills of 10.5~$\mu$s could be exploited,
which would require reducing the timing uncertainty to $\sim 1$~ns,
these figures would be improved significantly, to $M_{\nu QG1}\sim 5
\times 10^{7}$~GeV ($M_{\nu QG2} \sim 4 \times 10^{4}$~GeV). Using
events in the rock upstream of OPERA, and again assuming a time
resolution $\sim 1$~ns, the sensitivities to LV could
be further improved to $M_{\nu QG1} \simeq 4\times 10^8\ {\rm GeV}$
for the linear case and $M_{\nu QG2} \simeq 7 \times 10^{5}\ {\rm
GeV}$ for the quadratic case. 

\ack
Alexander Sakharov expresses his gratitude to the organizers of
DICRETE'08 for financial support, warm hospitality and the enthusiastic
scientific atmosphere of the meeting.

\end{document}